\documentclass[useAMS,usenatbib]{mn2e}

\usepackage {graphicx}
\usepackage {times}

\title[Spatially-Resolved Spectroscopy of an E+A (post-starburst) System with the Subaru telescope]{Spatially-Resolved Medium Resolution Spectroscopy of an Interacting E+A (post-starburst) System with the Subaru telescope\footnotemark[0]\thanks{Based  on data collected at Subaru Telescope, which is operated by the National Astronomical Observatory of Japan.}}
\author[Goto, Yagi,\& Yamauchi]{Tomotsugu Goto$^{1,2}$ %\footnotemark[1]
 \thanks{E-mail:tomo@ifa.hawaii.edu} \thanks{JSPS SPD Fellow}\thanks{Visiting Astronomer, Kitt Peak National Observatory, National Optical Astronomy Observatory, which is operated by the Association of Universities for Research in Astronomy (AURA) under cooperative agreement with the National Science Foundation.} , Masafumi Yagi$^{2}$, and  Chisato Yamauchi$^{3}$ %(author list and order is temporary. Suggestions welcome.)
\\
$^{1}$
Institute for Astronomy, University of Hawaii
2680 Woodlawn Drive, Honolulu, HI, 96822, USA\\
$^{2}$National Astronomical Observatory, 2-21-1 Osawa, Mitaka, Tokyo
181-8588,Japan\\
$^{3}$ Institute of Space and Astronautical Science,  Japan Aerospace Exploration Agency,
 3-1-1 Yoshinodai, Sagamihara, Kanagawa 229-8510, Japan\\
\\
}

\begin{document}
%--- DRAFTCOPY: COMMENT OUT IF NOT NEEDED -------
% Prints a large "DRAFT" diagonally across each page
% Does not show up in TeXview
% \typeout{Prints "DRAFT" on each page; does not show in TeXView}
% \special{!userdict begin /bop-hook{gsave 200 30 translate
% 65 rotate /Times-Roman findfont 216 scalefont setfont
% 0 0 moveto 0.90 setgray (DRAFT) show grestore}def end}
%%------------------------------------------------
\def\Hg{H$\gamma$}
\def\Hd{H$\delta$}

%\date{Accepted 2006 December 15. Received 1988 December 14; in original form 2006 March 17}
\date{\today; in original form 2008 April 28}

\pagerange{\pageref{firstpage}--\pageref{lastpage}} \pubyear{2008}

\maketitle

\label{firstpage}

\begin{abstract}
% E+A galaxies have been interpreted as post-starburst galaxies based on
% the presence of strong Balmer absorption lines combined with the
% absence of major emission lines ([OII] nor H$\alpha$). As a population
% of galaxies in the midst of the transformation, E+A galaxies has been
% subject to an intense research activity. %It has been, however,
% difficult to investigate E+A galaxies statistically since E+A galaxies
% are an extremely rare population of galaxies ($<$1\% of all galaxies).

  We have performed a spatially-resolved medium resolution long-slit spectroscopy of 
 a nearby E+A (post-starburst) galaxy system, SDSSJ161330.18+510335.5, with the FOCAS spectrograph mounted on the Subaru telescope. 
 This E+A galaxy has an obvious companion galaxy 14kpc in front with the velocity difference of 61.8 km/s. Both galaxies have obviously disturbed morphology. Thus, this E+A system provides us with a perfect opportunity to investigate the relation between the post-starburst phenomena and galaxy-galaxy interaction.

 We have found that H$\delta$ equivalent width (EW) of the E+A galaxy is greater than 7\AA~ galaxy wide (8.5 kpc) with no significant spatial variation. The E+A galaxy have a weak [OIII] emission (EW$\sim$1\AA) by $\sim$2.6 kpc offset from the peak of the Balmer absorption lines. 
 We detected a rotational velocity in the companion galaxy of $>$175km/s.  The progenitor of the companion may have been a rotationally-supported, but yet passive S0 galaxy. 
 %The progenitor of the E+A may have been pressure-supported but yet (puzzlingly) gas rich.
 We did not detect significant rotation on the E+A galaxy.
Metallicity estimate based on the $r-H$ colour suggests $Z=0.008$ and 0.02, for the E+A and the companion galaxies, respectively.
 Assuming these metallicity estimates, the age of the E+A galaxy after quenching the star formation is estimated to be 100-500Myr, with its centre having slightly younger stellar population.
 The companion galaxy is estimated to have older stellar population of $>$2 gigayear of age with no significant spatial variation.
 
 These findings are inconsistent with a simple picture where the dynamical interaction creates infall of the gas reservoir that causes the central starburst/post-starburst. Instead, our results present an important example where the galaxy-galaxy interaction can trigger a galaxy-wide post-starburst phenomena.

\end{abstract}


\section{Introduction}

\begin{figure*}
\begin{center}
\includegraphics[scale=0.6]{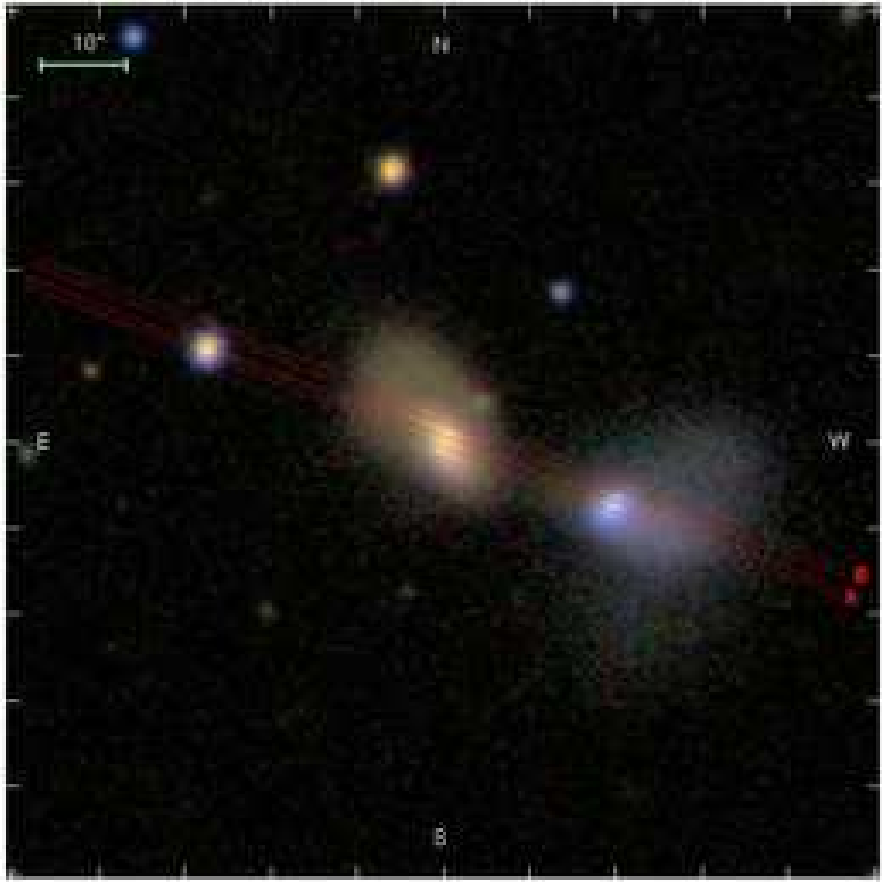}
\end{center}
\caption{The SDSS $g,r,i$-composite image of the J1613+5103. The long-slit positions are overlayed. 
 The E+A galaxy is to the right (west), with bluer colour. The companion galaxy is to the left (east).
}\label{fig:fchart}
\end{figure*}

\begin{figure*}
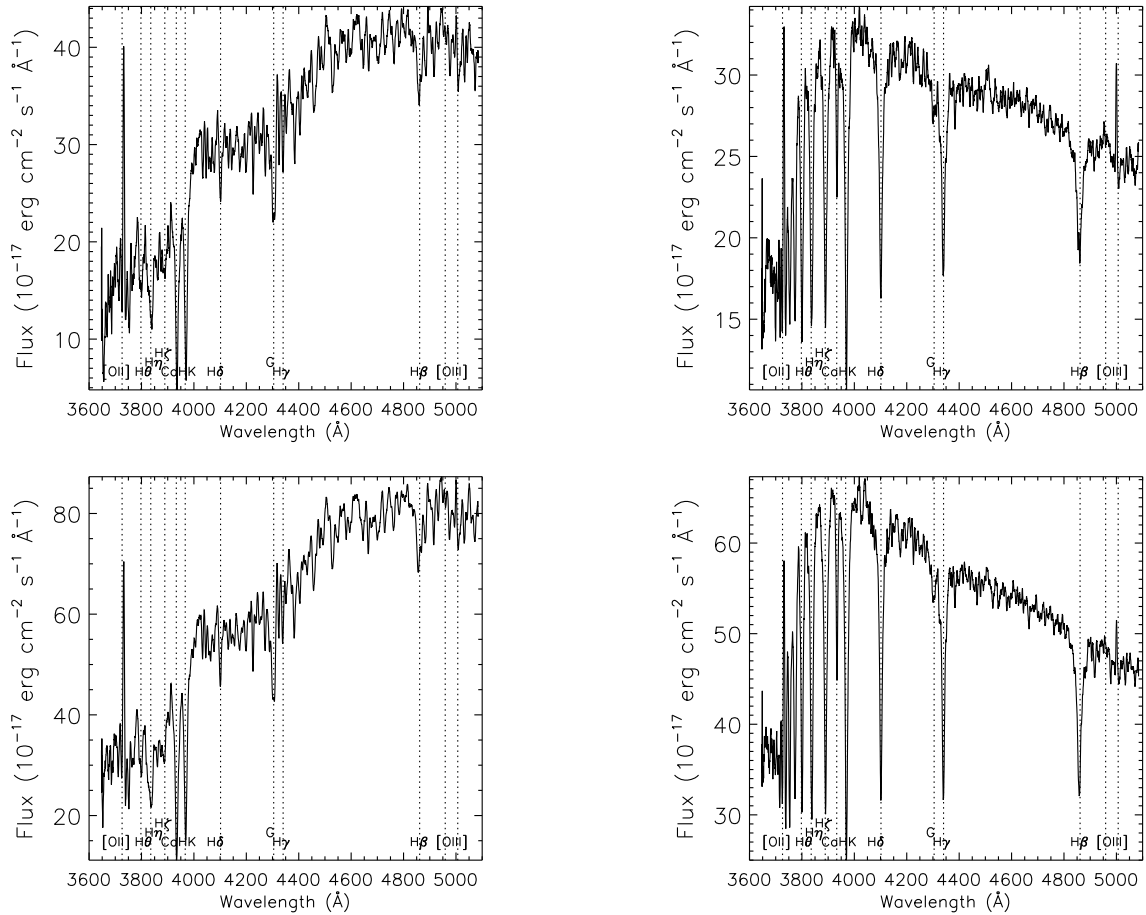

\begin{center}
\includegraphics[scale=0.49]{080112_J1613.ps_pages2}
\includegraphics[scale=0.49]{080112_J1613.ps_pages4}
\includegraphics[scale=0.49]{080112_J1613.ps_pages1}
\includegraphics[scale=0.49]{080112_J1613.ps_pages3}
\end{center}
\caption{Spectra of the entire galaxy. Right panels are for the E+A galaxy. Left panels are for the companion galaxy.
Bottom panels are from the core. Top panels are through the slit shifted by 2'' to the north.  The spectra are shifted to the rest-frame and smoothed using a 5-pixel box.  
}\label{fig:J1613_whole}
\end{figure*}

 It has long been known that galaxies have varieties of morphology; 
 Edwin Hubble was the first to classify galaxies into the so-called Hubble tuning fork, which mainly consists of elliptical, lenticular, spiral and barred spirals \citep{1926ApJ....64..321H}. Because of similarity and gradual change of galaxy properties along the Hubble's tuning fork, people speculated this might be an evolutionary sequence, and extensively studied the galaxy evolutionary sequence since the dawn of the extragalactic astronomy \citep[e.g.,][]{1994ARA&A..32..115R}.
 However, it is still not very well understood what physical mechanism drives the galaxy evolution, creating the diverse galaxy properties such as in the  Hubble tuning fork.
As a galaxy in a transition phase, one of the key populations in elucidating this galaxy evolution is so-called E+A galaxies.

 Galaxies with strong Balmer absorption lines without any emission in [OII] nor H$\alpha$ are called E+A galaxies. 
 The existence of strong Balmer absorption lines shows 
 that E+A galaxies have experienced starburst recently \citep[within a gigayear;][]{2004A&A...427..125G}.  However, these galaxies do not show any sign of on-going star formation as non-detection in the [OII] emission line indicates.  
 Therefore, E+A galaxies have been interpreted as post-starburst galaxies,
 that is,  a galaxy which truncated starburst suddenly \citep{1983ApJ...270....7D,1992ApJS...78....1D,1987MNRAS.229..423C,1988MNRAS.230..249M,1990ApJ...350..585N,1991ApJ...381...33F,1996ApJ...471..694A}. Recent study found that E+A galaxies have $\alpha$-element excess \citep{2007MNRAS.377.1222G}, which also supported the post-starburst interpretation of E+A galaxies. 
 However, the reason why they started starburst, and why they abruptly stopped starburst remains one of the mysteries in the galaxy evolution. Since a post-starburst is an important stage of the overall galaxy evolution in the Universe, it is important to investigate the physical origin of the E+A galaxies.

  At first, E+A galaxies are found in cluster regions, especially at higher redshift \citep{1985MNRAS.212..687S,1986ApJ...304L...5L,1987MNRAS.229..423C,1988MNRAS.235..827B,1991ApJ...381...33F,1995A&A...297...61B,1996MNRAS.279....1B,1998ApJ...498..195F,1998ApJ...507...84M,1998ApJ...497..188C,1999ApJS..122...51D,1999ApJ...518..576P,2003ApJ...599..865T,2004ApJ...617..867D,2004ApJ...609..683T}. Therefore a  cluster specific phenomenon such as the ram-pressure stripping was thought to be responsible for the  violent star formation history of E+A galaxies \citep{1951ApJ...113..413S,1972ApJ...176....1G,1980ApJ...241..928F,1981ApJ...245..805K,1983ApJ...270....7D,1999MNRAS.308..947A,1999PASJ...51L...1F,2000Sci...288.1617Q,2003ApJ...584..190F,2003ApJ...596L..13B,2004PASJ...56..621F}.
 
 However, \citet{2004MNRAS.355..713B} found that low redshift
E+A galaxies are located predominantly in the field environment, suggesting that a physical mechanism that works in the field region is at least partly responsible for these E+A galaxies. 
Recently, \citet{2005MNRAS.357..937G} has shown that E+A galaxies have more close companion galaxies than average galaxies, providing a statistical evidence that the dynamical merger/interaction could be the physical origin of field E+A galaxies. Dynamically disturbed morphologies of E+A galaxies also support this scenario \citep{2006astro.ph.12053L,2005MNRAS.359.1557Y}.

 To understand the origin of E+As is cluster-related or merger-driven (or both), independent evidence on the origin of E+A galaxies has been waited.
  Previous work mentioned above has been focused on the investigation of the global/external properties of E+A galaxies, such as the environment of E+A galaxies \citep{2005MNRAS.357..937G}, and the integrated spectra of the E+A galaxies \citep{2004ApJ...617..867D}.
 However, if the physical origin of E+A galaxies is merger/interaction or gas-stripping, these mechanisms should leave traces to the spatial distribution of stellar-populations within each E+A galaxy. For example, a centrally-concentrated post-starburst region is  expected to be found in case of the merger/interaction origin \citep[e.g.,][]{1992ARA&A..30..705B}. In contrast, the gas-stripping would create a more uniform, galaxy-wide post-starburst region. Thus, the spatial-distribution of the post-starburst region inside the E+A galaxy contains important and independent clues on the physical origin of E+A galaxies \citep[e.g.,][]{2005MNRAS.359.1421P,2005ApJ...622..260S}.
 In this series of papers \citep{2006AJ....131.2050Y,YGH,goto_3D}, we try to obtain such independent hints on the origin of E+A galaxies by revealing internal structure of E+A galaxies. In this paper, we focus on a spatially-resolved long-slit scan spectroscopy of a nearby E+A system J1613+5103.
 This E+A galaxy previously studied by \citet{YGH} has an obvious companion galaxy in front (Fig.\ref{fig:fchart}), and thus, provides us with a perfect opportunity to reveal the relation between the post-starburst phenomena and galaxy-galaxy interaction. 
 Color gradient of the E+A galaxy is studied in detail by \citet{2005MNRAS.359.1557Y}.
 Compared to our own previous work \citep{2006AJ....131.2050Y,YGH,goto_3D}, this time we obtained E+A spectra with higher resolution and better signal-to-noise taking advantage of the Subaru telescope.

  Unless otherwise stated, we adopt the WMAP cosmology: $(h,\Omega_m,\Omega_L) = (0.7,0.3,0.7)$ \citep{2008arXiv0803.0547K}.

\begin{figure*}
\begin{center}
\includegraphics[scale=0.39]{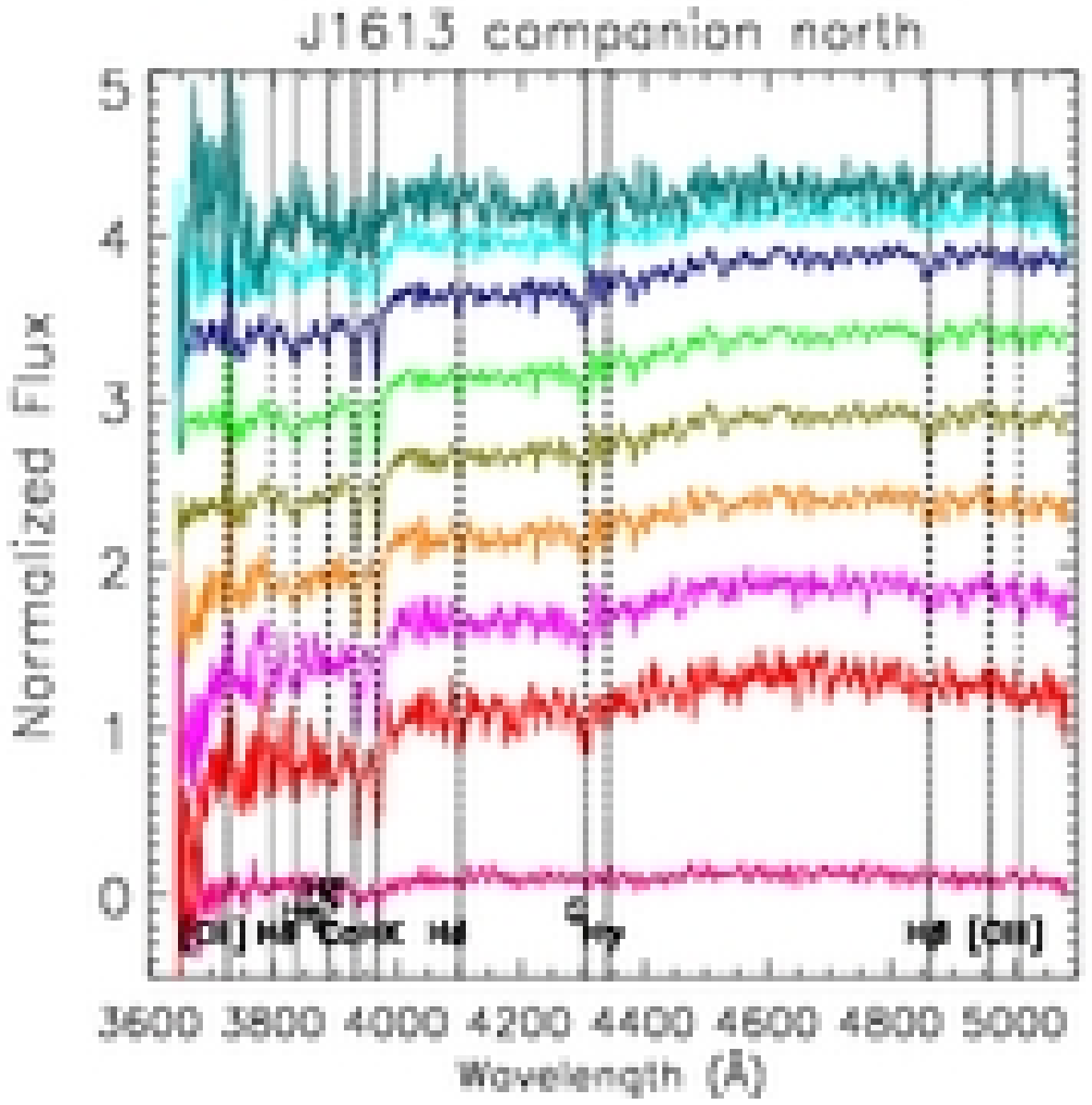}
\includegraphics[scale=0.39]{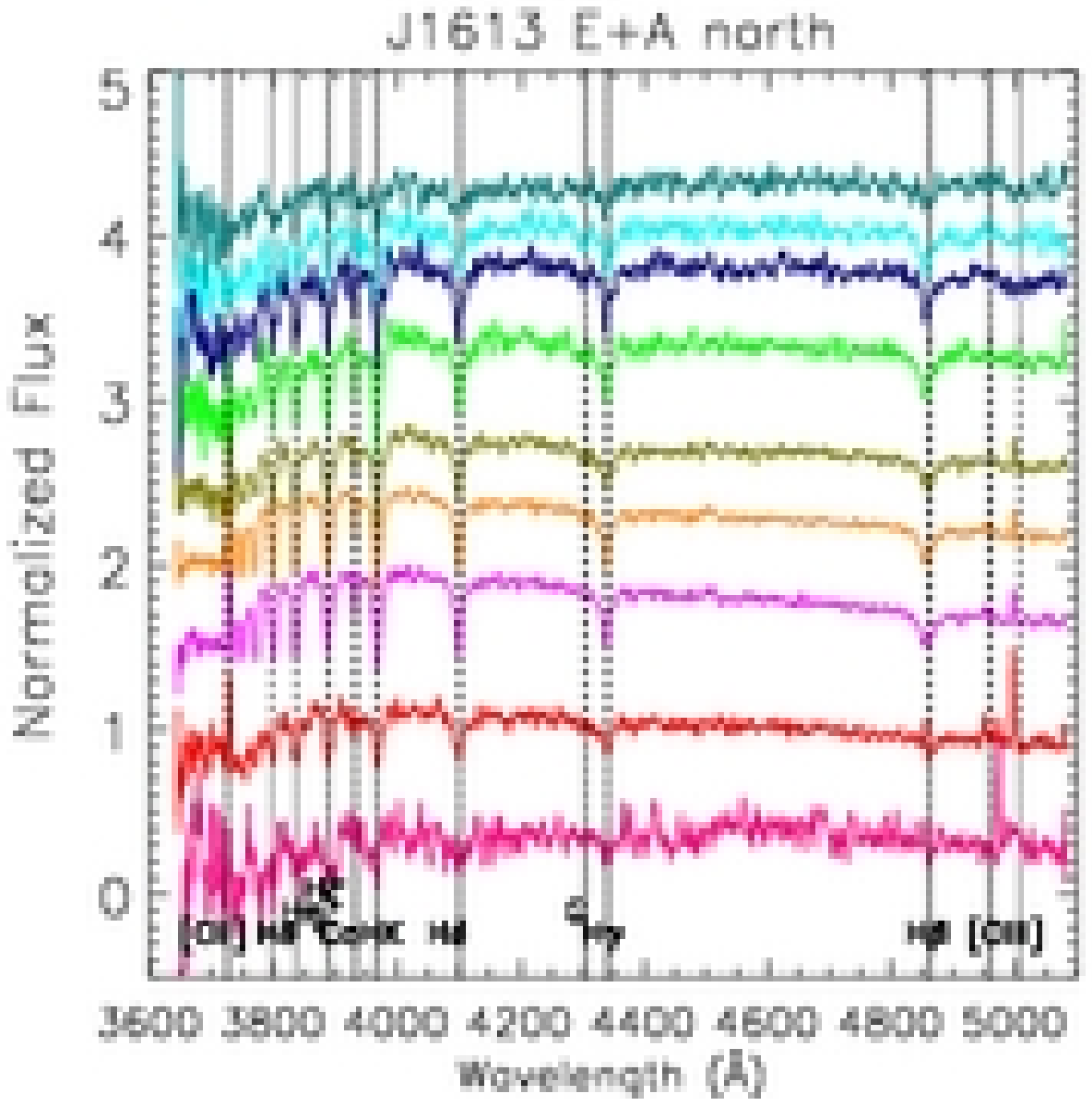}
\includegraphics[scale=0.39]{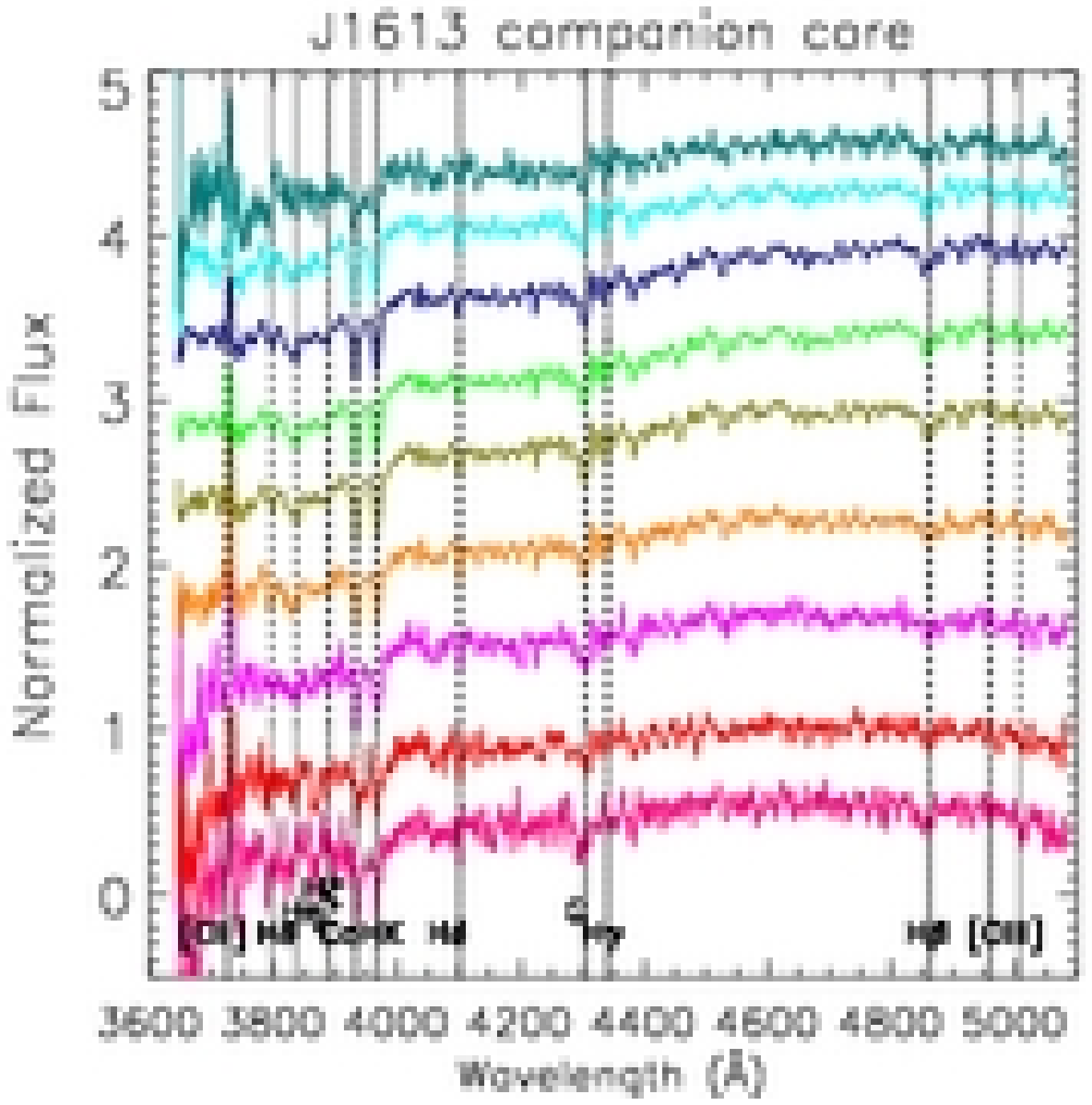}
\includegraphics[scale=0.39]{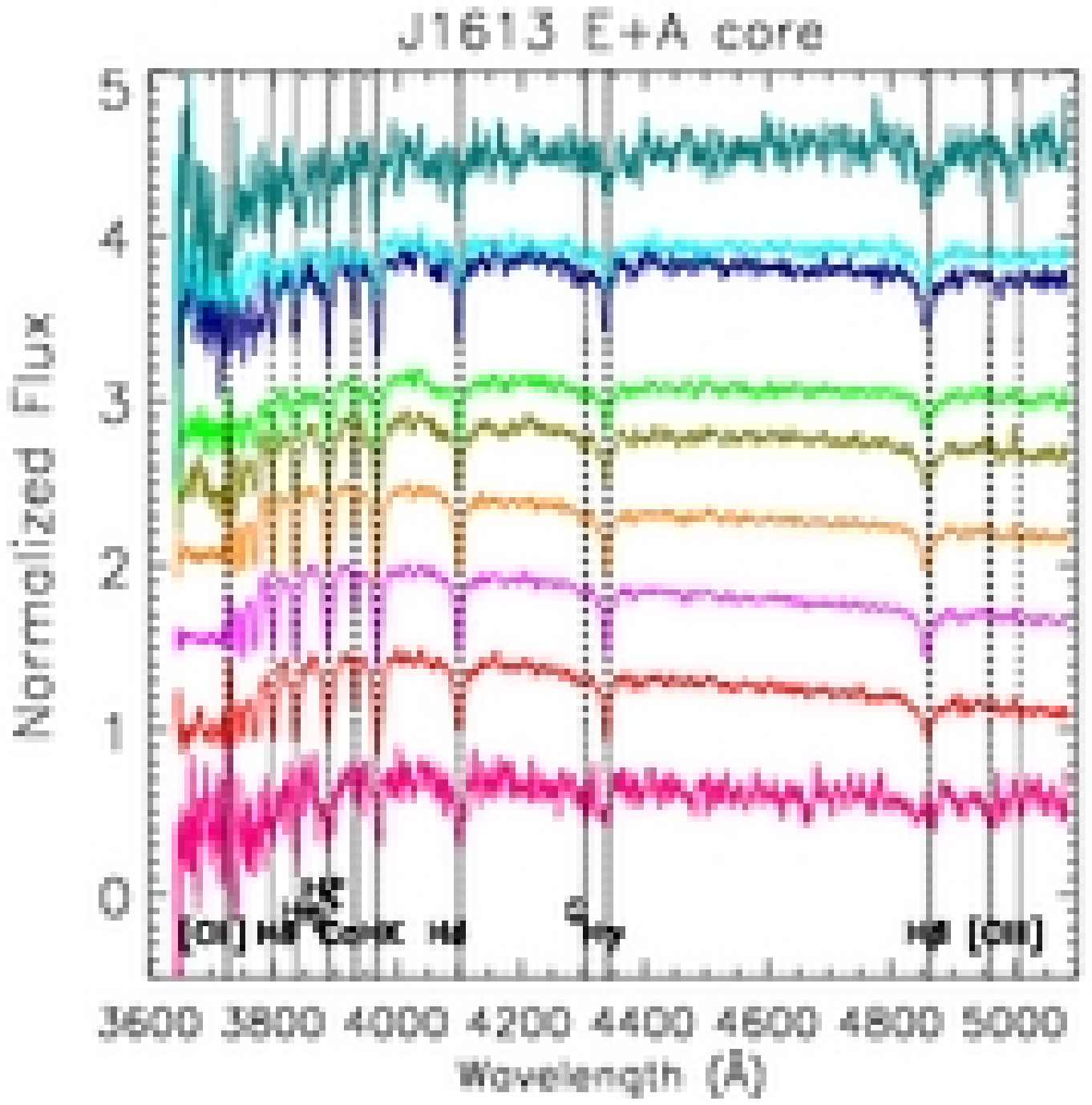}
\end{center}
\caption{
Spectra in each aperture.
 Right panels are for the E+A galaxy. Left panels are for the companion galaxy.
 Bottom panels are from the core. Top panels are shifted by 2'' to the north.  
 Each spectrum is spatially-divided into 9 equally-spaced bins. 
 In each panel, these 9 spectra are normalized by the peak flux and aligned in the vertical direction (thus, the flux scale is arbitrary). 
 The spectra are shifted to the rest-frame and smoothed using a 5-pixel box.  
}\label{fig:J1613_ap9}
\end{figure*}

\section{Sample Selection}\label{sample}

\begin{table*}
% \centering
 \begin{minipage}{180mm}
  \caption{Target properties}\label{tab:targets}
  \begin{tabular}{@{}llrrcccc@{}}
  \hline
%   Object     &            & \multicolumn{4}{c}{Flux density (Jy)%}
   Object & Redshift & $g_{AB}$ & $r_{AB}$ & Petro Rad in $r$ ('')  &   exposure time (min) & Observing date (HST) & $M_r$\\
%        &  &  &  &  &  & group & (d) & curve \\
%        &  &  &  &  &  &       &     & type  \\
 \hline
SDSSJ161330.18+510335.5 (E+A)& 0.0341$\pm$0.0009 & 15.86 & 15.38 &  19.6 &  50 (core), 60 (north) & July, 29, 2005 &     -20.5 \\
SDSSJ161332.23+510342.9 (companion)&  0.0339$\pm$0.001 &  15.89 & 15.09 & 4.6 &  50 (core), 60 (north)  & July, 29, 2005 &      -20.7 \\
\hline
\end{tabular}
\end{minipage}
\end{table*}

%  Strong Balmer absorption lines such as H$\delta$, H$\gamma$ and H$\beta$ show that the galaxy spectrum is dominated by A-type stars with the absence of massive O,B-type stars. Since the lifetime of A-stars is about one gigayear, these galaxies have experienced starburst within the last one gigayear and stopped it by now, i.e., they are in the post-starburst phase \citep{1983ApJ...270....7D,1999ApJ...518..576P,2004A&A...427..125G}. %By looking for galaxies with strong Balmer absorption, we can select post-starburst galaxies. 

 We selected our target E+A galaxy, SDSSJ161330.18+510335.5, from \citet{2005MNRAS.357..937G,goto_DR6}, which presented a catalog of 564 E+A galaxies based on the $\sim$670,000 galaxy spectra of the Sloan Digital Sky Survey \citep{2006ApJS..162...38A}. 

 The galaxy has extremely strong Balmer absorption lines with the H$\delta$ equivalent width (EW) of 7.37\AA\footnote{Absorption lines have a positive sign throughout this paper.}, without significant emission in [OII] nor H$\alpha$ within the three arcsec diameter of the SDSS fiber spectrograph. These are representative characters of an E+A galaxy and indicate that this galaxy is in the post-starburst phase.

 The E+A galaxy has an apparent companion galaxy, SDSSJ161332.23+510342.9, in front. 
Note that the companion galaxy is perhaps as massive as the E+A galaxy (see Section\ref{sec:rotation}), even though we call it a ``companion''.   
Previously, \citet{YGH} confirmed that the companion galaxy is at the same redshift, i.e., these two galaxies are dynamically-interacting.
 Along with the large apparent size of $\sim$20 arcsec on the sky, this E+A system is a perfect target for spatially-resolved spectroscopy.
 
 We present the basic properties of the E+A and companion galaxies in table \ref{tab:targets}, where measured quantities such as positions, redshift, magnitudes in $g$ and $r$, and Petrosian radius are taken from the SDSS catalog. %Here, $R_{90}$ is the radius within which 90\% of the $r$-band Petrosian flux is contained. 
 Magnitudes are de-reddened (for the Galactic extinction) Petrosian AB-magnitude in $g$ and $r$-band. 
%The physical scale based on the WMAP cosmology is shown in the unit of kpc/arcsec for convenience. 

\section{Observation}
\label{Observation}

 Observations were carried out on July 29, 2005 with the FOCAS spectrograph \citep[The Faint Object Camera and Spectrograph;][]{2002PASJ...54..819K} on board the Subaru telescope.
 We performed a long-slit spectroscopy with the slitwidth of 1.0'' on the position A (core) and B (north) of Fig.\ref{fig:fchart}. Both slits cover the E+A galaxy and the companion galaxy simultaneously. We used the {\ttfamily VPH350} grism, which provides us with the wavelength coverage of 3800-5250\AA, with the resolution of $R\sim$1600. We used 3$\times$1 binning mode, giving the pixel size of 0.311''/pixel in a spatial direction. The seeing was $\sim$1.6'' based on a bright star we simultaneously placed on the slit. The exposure time was 600+1200$\times$2 seconds for the slit A (centre), and 1200$\times$3 seconds for the slit B (north). We moved the target along the slit $\pm$2 arcsec between the exposures to remove systematic effects.

 The data reduction was performed with the standard procedure with the IRAF; The overscan subtraction, distortion correction, and the flat-fielding were performed.  We used a Th-Ar lamp for the wavelength calibration. Then, frames were combined after the spacial shift were accounted for. Background subtraction and the extraction of the spectra are carried out in a standard manner. 
A standard star HZ44 observed with the 2.0'' slit was used for the flux calibration. 
In Fig.\ref{fig:J1613_whole}, we show spectra of the two galaxies obtained from each slit. It is immediately noticed that the E+A galaxy has strong Balmer absorption lines along with an A-star like continuum.
 The spectra of the companion galaxy are dominated by old stars with deep Ca H\&K absorptions. 
 Both galaxies have weak emissions in [OII] and possibly in [OIII], which will be discussed later.
 Measured redshifts based on [OII], CaH\&K lines are 0.0341$\pm$0.0009 and 0.0339$\pm$0.001 for the E+A and companion, respectively. These values are consistent with the redshift measured by the SDSS (0.034$\pm$0.00012) and by \citet{YGH}. The relative velocity is 61.8 km/s. One arcsec corresponds to 0.66 kpc at this redshift.
 
 One of the main purpose of this observation was to carry out spatially-resolved analysis. To this purpose, we have divided the 20.3-25.5 arcsec of aperture into 9 equally spaced bins for each spectrum. And thus, the size of each bin (2.3-2.8'') is larger than the PSF size of 1.6''. 
In Fig. \ref{fig:J1613_ap9}, we show spectrum of each bin for both galaxies.

Using these spectra, we measured equivalent widths of [OII], H$\delta$, H$\gamma$, H$\beta$ and [OIII] lines using the flux summing technique in \citet{2003PASJ...55..771G}. The wavelength ranges used are presented in \citet{1997ApJS..111..377W,2002AJ....124.2453M}.
The strength of the 4000 \AA~ break (D4000) was measured using the definition in \citet{2002AJ....123..485S}.

\begin{figure*}
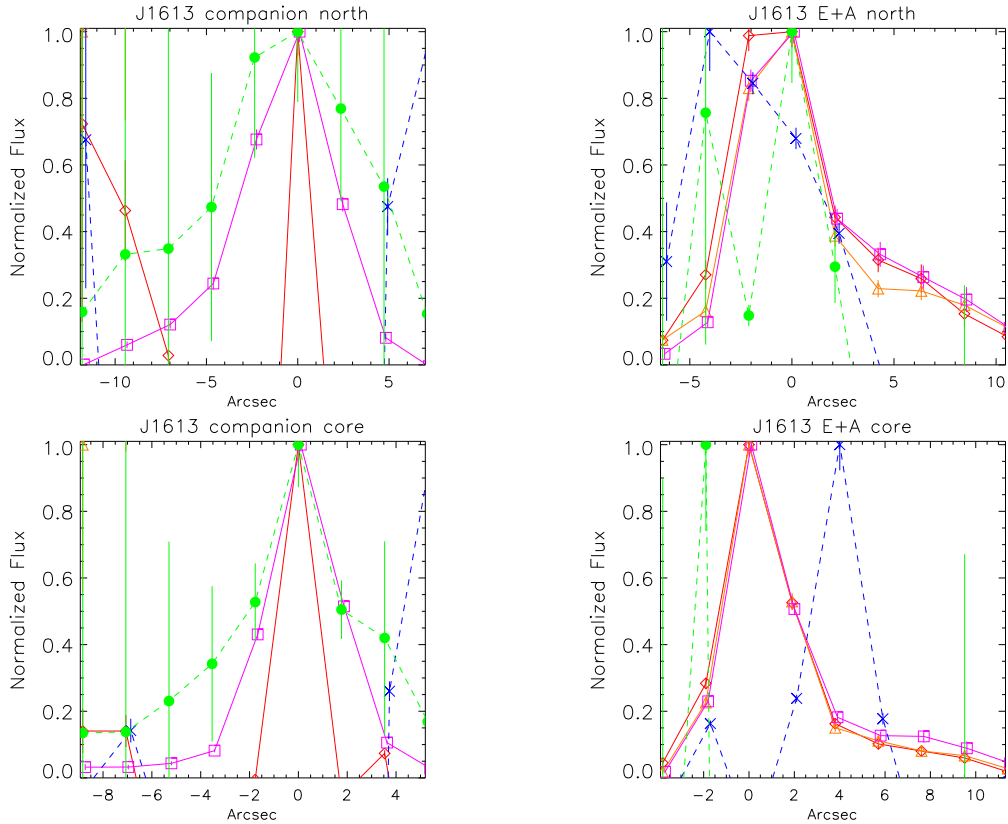

\begin{center}
\includegraphics[scale=0.43]{080125_J1613_ap9.ps_pages_flux4}
\includegraphics[scale=0.43]{080125_J1613_ap9.ps_pages_flux2}
\includegraphics[scale=0.43]{080125_J1613_ap9.ps_pages_flux3}
\includegraphics[scale=0.43]{080125_J1613_ap9.ps_pages_flux}
\end{center}
\caption{Normalized flux profiles:
 The right panels show the normalized flux profiles for the E+A core (lower) and north (upper) spectra. 
 The left panels show the normalized flux profiles for the companion core (lower) and north (upper) spectra. 
 The  diamond (red), triangle (orange), and square (pink) symbols are for  H$\delta$, H$\gamma$ and H$\beta$, respectively.
 These absorption lines are connected with the solid lines.
 The cross (blue) and filled (green) circle are for [OIII] and [OII] emission lines, which have positive sign in this plot.  The emission lines are connected with the dotted lines.
 The abscissa is shifted so that the H$\delta$ flux peaks at $X=0$. 
 One arcsec corresponds to 0.66 kpc at this redshift.
}\label{fig:flux}
\end{figure*}

\section{Results}\label{results}

\subsection{Line flux profile}

 In Fig.\ref{fig:flux}, we show normalized flux profiles of the measured lines. Profiles are normalized at a peak flux. 
 Fluxes for absorption lines are measured by integrating the amount of absorption under the fitted-continuum, i.e., they are pseudo-flux measured in a similar way to emission line fluxes.
The right two panels show the profile for the E+A galaxy. Absorption lines (H$\delta$,H$\gamma$,and H$\beta$) connected with the solid lines show a perfect agreement with each other. 
 Interestingly, [OIII] emission line peaks 4 arcsec away from the absorption peak on the bottom-right panel and -4 arcsec away on the top-right panel, possibly indicating that the current star formation is going on at a different location from the post-starburst region. [OII] emission is too noisy to locate the peak of the profile.
 
 For a companion galaxy on the left panels, only H$\beta$ (magenta squares) and [OII] (green filled circles) have enough signal-to-noise to measure profiles. The peaks of these two profiles agree with each other. The [OII] emission shows a slightly extended profile. %, indicating the remaining star formation occurs in a more extended region than that of the post-starburst. 

\subsection{Equivalent widths profile}

\begin{figure*}
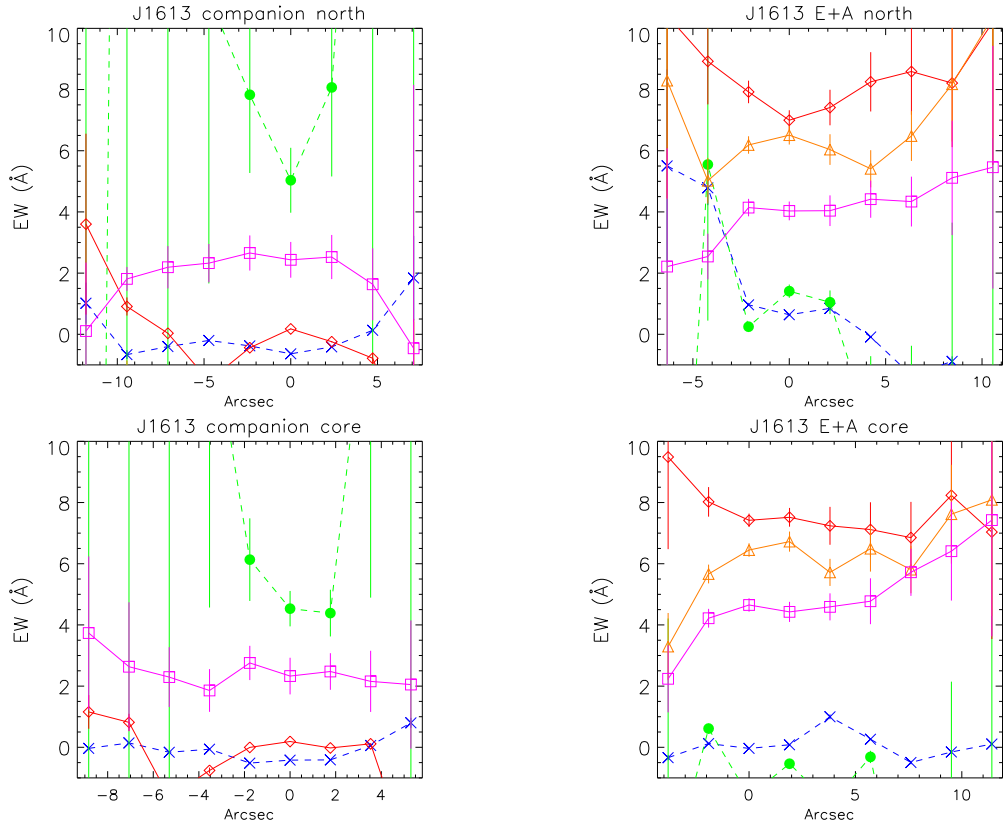

\begin{center}
\includegraphics[scale=0.43]{080204_J1613_ap9.ps_pages_ew4}
\includegraphics[scale=0.43]{080204_J1613_ap9.ps_pages_ew2}
\includegraphics[scale=0.43]{080204_J1613_ap9.ps_pages_ew3}
\includegraphics[scale=0.43]{080204_J1613_ap9.ps_pages_ew}
\end{center}
\caption{
Equivalent width (EW) profiles:
 The right panels show the EW profiles for the E+A core (lower) and north (upper) spectra. 
 The left panels show the EW profiles for the companion core (lower) and north (upper) spectra. 
  The  diamond (red), triangle (orange), and square (magenta) symbols are for  H$\delta$, H$\gamma$ and H$\beta$, respectively. These absorption lines are connected with the solid lines.
 The cross (blue) and filled (green) circle are for [OIII] and [OII] emission lines, which have positive sign in this plot. The emission lines are connected with the dotted lines.
  One arcsec corresponds to 0.66 kpc at this redshift.
}\label{fig:ew}
\end{figure*}

  In Fig.\ref{fig:ew}, we present EW distribution of three absorption lines (H$\delta$,H$\gamma$,and H$\beta$)  and two emission lines ([OII] and [OIII]) for the E+A/companion galaxies in the core/north spectra.
The right panels show EWs for the E+A galaxy. The bottom panel is from the core, the top panel is from the north spectra.
 The symbols are the same as the Fig.\ref{fig:flux}.
 Unlike the flux profile (Fig.\ref{fig:flux}), EWs are not affected by the luminosity profile of the galaxy, and thus, suitable to trace the efficiency in producing emission/absorption.
  In the right panel of Fig.\ref{fig:ew}, all the absorption lines have very large EW galaxy wide; H$\delta$ (red diamonds) EW is $>7$\AA~ galaxy wide in both core/north spectra ($>$13 arcsec wide or $>$8.5 kpc). There is no significant radial trend for H$\delta$. 
 H$\gamma$ (orange triangles) and H$\beta$ (magenta squares)EWs stay at $\sim$6 and 4 \AA, respectively, for the entire galaxy except for the low S/N regions at the edge. 
 Emission lines have very weak EWs: [OIII] EWs are 1\AA~ or less for everywhere; [OII] EWs are also small. Although in Fig. \ref{fig:J1613_ap9}, we see a spectrum with emission lines, these emission lines are weak as in Fig.\ref{fig:ew}. Therefore, we can still call this galaxy as an E+A galaxy.

 For a companion galaxy, spectra are shown in the left panels of Fig.\ref{fig:J1613_ap9}. The bottom-left panel is for the core spectrum, the top-left panel is for the north spectrum. 
 For this companion galaxy, H$\delta$ and H$\gamma$ absorption lines are weak, indicating the stellar population is old.  The only line securely detected is H$\beta$, which has a flat profile with $\sim$2\AA~ of EWs. There is no significant radial trend. [OII] line is detected with 4-5\AA~ around the galaxy centre. The [OII] emission is stronger for the companion than for the E+A.

\subsection{Rotation}\label{sec:rotation}

 With the resolution of $R\sim1600$, we have a chance to measure rotational velocity of galaxies.
We fit a Gaussian to the H$\delta$ (absorption) line, to measure the red/blue shifted central wavelength. 
 In other occasions, emission lines such as [OII] or [OIII] are preferred for this exercise. In our particular case, however, the H$\delta$ (absorption) line is much stronger than the emission lines, providing us with a better measurement of the rotation curve.% In the literature, absorption lines were used to measure rotation of early-type galaxies \citep[e.g.,][]{2008ApJS..175..462C}.
 We visually inspected the goodness of the Gaussian fit and only spectra where the line has sufficient signal-to-noise ratio are used. 
%We also checked that the rotation measurement with [OII] or [OIII] lines returned a similar results with smaller significance.

\begin{figure*}
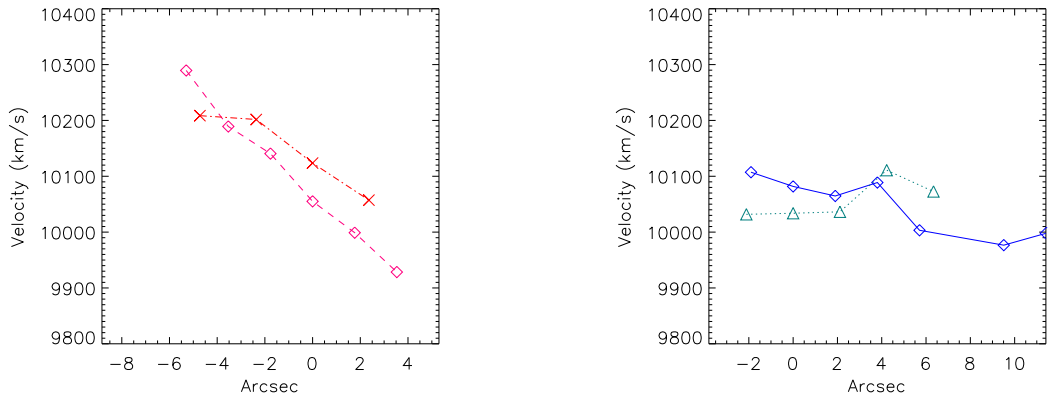

\begin{center}
\includegraphics[scale=0.45]{080423_J1613_ap9.ps_pages_hd_comp}
\includegraphics[scale=0.45]{080423_J1613_ap9.ps_pages_hd_ea}
\end{center}
\caption{Recession Velocity. 
 %We plot central wavelength obtained from Gaussian fit to [OII] (left panel) and [OIII] (right panel) emission lines for the E+A and companion galaxies.  
 We plot recession velocity of the H$\delta$ line.
 The diamonds and triangles in the right panel are for the E+A core/north spectra, respectively. The squares and crosses are for the companion galaxy's core/north spectra in the left panel. 
 We visually checked the gaussian fits, and only data points based on a reasonably good fit are shown.
 One arcsec corresponds to 0.66 kpc at this redshift.
}\label{fig:velocity}
\end{figure*}

In Fig.\ref{fig:velocity}, we show the recession velocity from the H$\delta$ line. In the right panel, the diamonds (blue) and triangles (green) are for the core/north E+A spectra. 
 There may be a slight sign of the rotation of $\sim$50km/s in the core spectra (blue diamonds). However, this is below the resolution of the spectra, and the trend can not be recognized in the north spectra (green triangles). Therefore, we do not claim a rotation of the core/north E+A spectra.

 The squares (magenta) and crosses (red) in the left panel of Fig.\ref{fig:velocity} are for the companion galaxy in the core/north.
 A clear rotation curve is found from the position of the $X=-6$ arcsec to $X=+3$ arcsec.
 %, with the velocity of $>$97.3 km/s. 
% The signal is weaker in the [OIII] line shown in the right panel due to much weaker emission in [OIII] than [OII] for the companion galaxy. %We, however, see $>$79.6 km/s of rotation. 
% In the bottom panel, we investigate the rotation using the H$\delta$ absorption line. 
 The rotation observed for the companion galaxy is greater than 175 km/s.

 It is interesting to find no significant rotation for a galaxy that has just experienced a starburst (E+A). This suggests following possibilities: (i) the galaxy-galaxy dynamical interaction destroyed the rotation of the galaxy (however the companion galaxy has the rotation); (ii) the progenitor of the E+A galaxy was not a rotationally-supported disk galaxy, although it possessed enough gas to create a starburst; or (iii) the rotation of the E+A galaxy is face-on although we do not see an obvious face-on disk in Fig.\ref{fig:fchart}.
 
 It is also puzzling that the companion galaxy whose spectra look like that of an elliptical galaxy has rotational velocity. One possibility is that the rotational velocity originates from the dynamical interaction, although no rotation was found for the E+A galaxy.
  Another possibility is that in Fig.\ref{fig:fchart}, the core of the companion galaxy is slightly extended toward the north-east direction, implying a possible position-angle of the former disk component. Thus, the companion galaxy may have been a rotationally-supported disk galaxy before the dynamical interaction started.

We also tried to measure the velocity dispersion for both the E+A and the companion galaxies. Our method is similar to that utilized by the SDSS spectroscopic pipeline; first we masked out possible emission lines, then the rest of the spectra are fit with the combination of eigen spectra with varying velocity dispersion. The measured velocity dispersions are 279$\pm$49,  224$\pm$31 km/s for the E+A and companion galaxies, respectively. The results suggest that galaxies have comparable mass, and thus, the system is experiencing an equal-mass (major) merger with the mass ratio close to 1.
The derived mass ratio is consistent with \citet{2001ApJ...547L..17B}'s prediction that requires mass ratio of $>0.3$ to create an E+A galaxy.

%xxx
%
%
%It has been suggested that the galaxy merger might create E+A galaxies \citep[e.g.,][]{2005MNRAS.357..937G}. 
%Having a close companion at the same redshift, this E+A galaxy, J1656, is a real example produced by the galaxy-galaxy merger with the upper-left companion. At the same time, J1656 presents an interesting example where the companion galaxy is not necessary an E+A galaxy, even though  it involves with the same galaxy merger. This example suggests that not only a merger but an additional condition is required to produce an E+A galaxy. 
% Only a few spectroscopic companions of E+A galaxies are known before. For example, a companion galaxy of an E+A galaxy in \citet{YGH} was a passive galaxy. More spectroscopic follow-up of E+A companion galaxies is needed to reveal what kind of galaxy merger can produce an E+A galaxy. 

\subsection{Metallicity estimate}\label{metal}

\begin{figure}
\begin{center}
\includegraphics[scale=0.3]{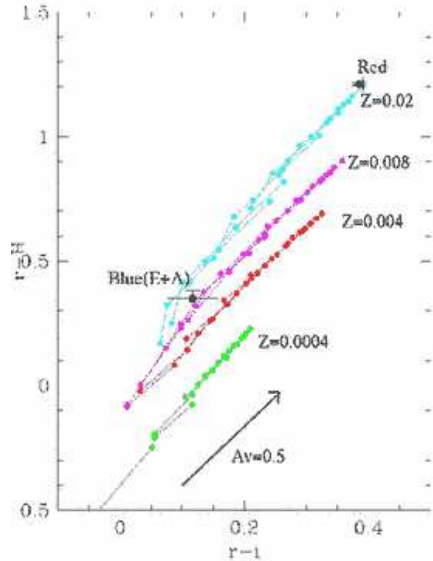}
\end{center}
\caption{
 The $r-i$ vs $r-H$ colour-colour diagram. The solid lines
with small dots are models based on \citet{2003MNRAS.344.1000B} at z=0.034
with varying metalicity. Colours of E+A and companion galaxy are
shown with filld circles. 
%The $r-i$ vs. $r-H$ colour-colour diagram. The solid lines with small points are models based on \citet{2003MNRAS.344.1000B} with varying metallicity. Colours of the E+A and companion galaxy are shown with the black dots.
 }\label{fig:nir}
\end{figure}

Next, we would like to estimate age and burst strength (stellar mass ratio of young population to old stellar population) of these two galaxies by comparing observational data with models.
To do this, however, we first need to estimate metallicity of these galaxies since there is a well-known fact that galaxy age and metallicity are degenerated in optical colour or spectra. This age-metallicity degeneracy can be lifted when you have near-infrared photometry in hand; optical-near-infrared colour such as $V-K$ is a measure of the temperature of its red giant branch. The temperature is strongly dependent on metallicity but has little sensitivity to age \citep{2002A&A...391..453P,2006ARA&A..44..193B,2007ApJ...669..982C}. 
In Fig.\ref{fig:nir}, we plot $r-H$ colour against $r-i$ colour. 
Optical colours are from the SDSS. The $H$-band image was taken with the SQUIID NIR camera on board the KPNO2m telescope on the night of June 4th, 2006. A complete analysis of these data will be presented elsewhere. The seeing sizes were similar among $r,i$ and $H$-bands. Fluxes are measured in a fixed aperture of 3'' at the centre. 

 In Fig.\ref{fig:nir}, Overplotted lines are the SED models;
 We constructed model SEDs with
GALAXEV \citep[hereafter BC03]{2003MNRAS.344.1000B}, using 
the extinction law by \citet{1989ApJ...345..245C}, and Salpeter IMF. 
Metallicities are changed in the range of $0.0004\leq Z\leq 0.02$. 
The model galaxies evolve for 10Gyr with an exponentially
decreasing star-formation rate ($\tau$=1 Gyr), and then a secondary instantaneous burst occurs;
 The dots represent 100, 200, 300, 500 Myrs and 1Gyr of time after the instantaneous burst.

This $(r-i)$ vs $(r-H)$ diagram is advantageous in determining the
metallicity of galaxies when a spectrum does not cover Fe or Mg
lines. Three colours ($r,i,H$) are carefully chosen so that age,
burst-strength, and dust-extinction are degenerate in one (straight)
direction perpendicular to the metallicity variation. Only possible
alternative is $r-z-K_s$ colour among general broad-band optical-NIR
filters. In other choices of colours, the models curve and overlaps with
each other.  In fact in Fig.\ref{fig:nir}, age vectors (lower-left to
upper-right in the Figure) are almost perpendicular to the metallicity
change.  The models with different burst-strengths (we checked 50, 30,
10, 5, 3 and 1\%) are almost overlapped with each other, i.e., no
degeneracy with the metallicity.  Moreover, the dust reddening vector
(shown with an arrow in the Figure) is almost parallel to the aging
vector in this diagram. Therefore, the $(r-i)$ vs $(r-H)$  diagram provides us with  
a useful tool to measure metallicity of nearby poststarburst galaxies whose young population is 100M-1Gyr.

% This $r-i-H$ diagram is advantageous in determining the metallicity of galaxies when a spectrum does not cover Fe or Mg lines. Three colours ($r,i,H$) are carefully chosen so that age, burst-strength, and dust-extinction are degenerate in one (straight) line perpendicular to the metallicity variation (Only possible alternative is $r-z-K$ colour among general broad-band optical-NIR filters. In other choices of colours, the models curve and overlaps with each other).
%In fact in Fig.\ref{fig:nir},  age vectors (lower-left to upper-right in the Figure) are almost perpendicular to the metallicity change.
%  The models with different burst-strengths (we checked 50, 30, 10, 5, 3 and 1\%)
% are almost overlapped with each other, i.e., no degeneracy with the metallicity.
%Moreover, the dust reddening vector (shown with an arrow in the Figure) is almost parallel to 
%the aging vector in this diagram. Therefore, the $r-i-H$ diagram is a good indicator of galaxy metallicity.
% 

 Measured colours of the E+A galaxy and the companion galaxy are shown with the black dots in Fig.\ref{fig:nir}.
The E+A and companion galaxies are better described with the metallicity of $Z=0.008$ (i.e., 0.4$\times$solar metallicity) and $Z=0.02$, respectively. We will assume  $Z=0.008$ in the next section to estimate the E+A's age. 

In this paper, these colour based metallicity measurements are merely meant to aid the age determination in the following section. A more careful treatment such as high S/N spectroscopy covering Fe and Mg lines is needed if one wishes to obtain physical implications based on the metallicity measurement.

\subsection{Age estimate}

\begin{figure}
\begin{center}
\includegraphics[scale=0.6]{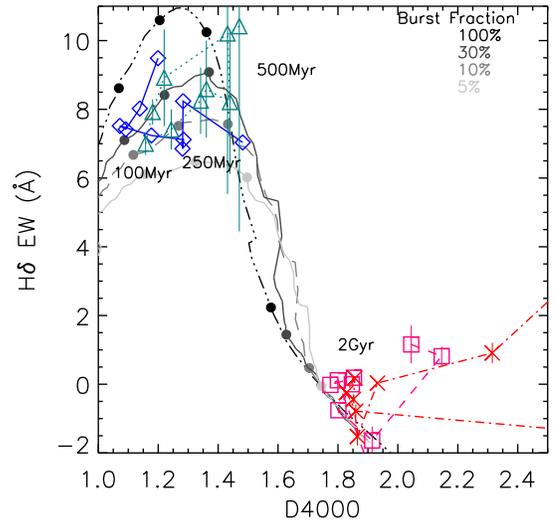}
\end{center}
\caption{
H$\delta$ EW is plotted against D4000. 
The diamonds and triangles are for the E+A core/north spectra, respectively.  The squares and crosses are for the companion galaxy's core/north spectra. 
Gray lines are population synthesis models from \citet{2003MNRAS.344.1000B} with 5-100\% delta burst population added to the 10G-year-old exponentially-decaying ($\tau$=1Gyr) underlying stellar population. Salpeter IMF and metallicity of $Z=0.008$ (See Section \ref{metal}) are assumed. On the models, burst ages of 0.1, 0.25, 0.5 and 2 Gyr are marked with the filled circles. See \citet{YGH} for more details of the models.
 }\label{fig:d4000}
\end{figure}

In Fig.\ref{fig:d4000}, we plot H$\delta$ EW against the strength of the 4000\AA~ break (D4000).
This diagram has been used as a good tool to estimate age, stellar-mass, and star-formation-history of galaxies \citep[e.g.,][]{2003PhDT.........2G,2003MNRAS.341...33K,YGH}. Symbols for data points are the same as in Fig.\ref{fig:velocity}. 
We overplot population synthesis models of \citet{2003MNRAS.344.1000B} with Salpeter IMF and metallicity of $Z=0.008$ (See Section \ref{metal}). We have checked that a small change in metallicity  $Z=0.008-0.02$ does not affect the discussion below.
Over the 10-Gyr-old stellar population with an exponentially-decaying star formation rate ($\tau$=1Gyr), we added 5,10, and 50\% (in terms of stellar mass) of instantaneous starburst population. The 100\% burst in the Figure is a pure burst with no underlying old stellar population. In each model, we marked 0.1, 0.25, 0.5, and 2 Gyrs of ages (after the instantaneous starburst) with the filled circles. These are the same models used in \citet{YGH}.

 By comparing models with data, E+A galaxies are most consistent with the 30\% burst model with the age of 250 Myr although the burst strengths of 5-30\% with the age of 100-500 Myrs are allowed within the error. 
%ここ重要！
 It seems that the core of the E+A galaxy has slightly yonger age ($\sim$100 Myr) than outskirts ($\sim$500 Myr). This trend is more consistent with the age difference than the difference in burst fraction, suggesting the galaxy core had experienced starburst more recently.
 One end facing to the companion seems to have a larger burst fraction ($>30\%$) than the other end (10\%).
 In any case, the comparison suggests that the E+A galaxy is in its very young phase ($\sim$250 Myr) of the post-starburst.
 On the other hand, the companion galaxy is consistent with much larger age of $>$2 Gyr. Although the companion galaxy has a weak [OII] emission, this indicates that major star formation activity of the companion galaxy ceased well-in-advance to the current galaxy merger/interaction. 

\begin{figure}
\begin{center}
\includegraphics[scale=0.6]{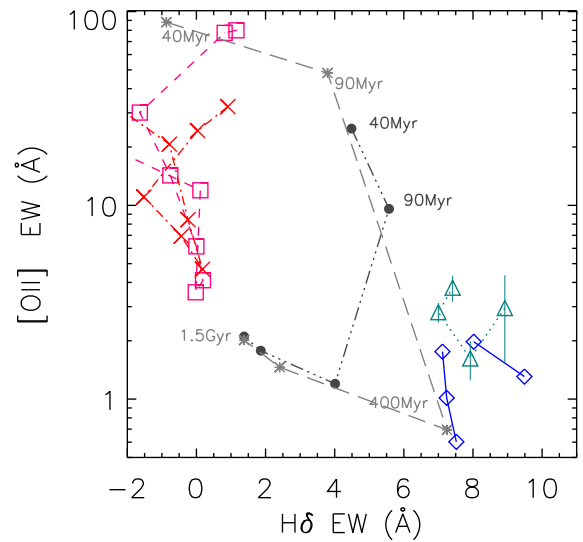}
\end{center}
\caption{
H$\delta$ EW is plotted against [OII] EW. 
The diamonds and triangles are for the E+A core/north spectra, respectively.  The squares and crosses are for the companion galaxy's core/north spectra.  Only data points with positive [OII] EW (i.e., in emission) are used. 
Gray lines are major-merger models from \citet{2001ApJ...547L..17B} with the ages (after the burst) of 40Myr-1.5Gyrs. Asterisks and filled-circles are for dust-free and dusty-starburst models, respectively. 
 }\label{fig:hd_oii}
\end{figure}

 In Fig.\ref{fig:hd_oii}, we compare [OII] EWs and H$\delta$ EWs. Symbols are the same as in previous figure.
 Overplotted models are from \citet{2001ApJ...547L..17B}. Their dusty-starburst and dust-free merger models of equal mass are with filled-circles and asterisks, respectively. We marked the ages (after the burst) of 40Myr-1.5Gyrs.
The E+A galaxy is consistent with $\sim$400 Myr of age of the dusty-starburst model after the starburst.
 The data points of the companion galaxies (magenta squares and red crosses)  deviate from the models, and cannot be explained within the model assumption of an equal mass merger of gas rich disks. Perhaps, the J1613+5103 is a wet-dry merger system, where the dry galaxy (companion) had a little remaining gas to induce star formation.

%-------------------------------------------------------------------------------------------------------------------

\section{Discussion}
\label{discussion}
% \begin{itemize}
%  \item Comparison to the previous work. Yagi et al.
%  \item Physical implication. Comparison to Pracey et al. 
%  \item rotational velocity for the companion? Puzzling.
%  \item Mass ratio. major merger, or minor merger?
% \end{itemize}

\subsection{Comparison to previous work}
 In this subsection, we summarize previous work on spatially-resolved spectroscopy of E+A galaxies, and compare them to our results.

  \citet{1996AJ....111...78C} obtained long-slit spectra of E+A galaxies in the Coma cluster and showed that starburst signatures are prominent in the central core and are spatially extended. % Cardwell
  Similarly, \citet{2001ApJ...557..150N} found that  the young stellar populations are more centrally concentrated than the older populations, but they are not confined to the galaxy core (radius $<$1 kpc).  %Norton
 Recently, \citet{2006AJ....131.2050Y} performed spatially-resolved spectroscopy of three E+A galaxies using the APO3.5m telescope. They found that the post-starburst region is as extended as the old population with no significant age gradient along the radius, concluding that the post-starburst in the E+As is a galaxy-wide phenomenon. 
%They found that H$\delta$ EWs were the largest at the galaxy centre although the strong H$\delta$ was significantly extended toward outside of the galaxies ($>4$kpc). %Yagi APO
%  Later,  \citet{YGH} observed a nearby E+A galaxy with a red companion galaxy at the same redshift (z=0.033) with the FOCAS long-slit spectrograph on the Subaru telescope. The spatially-resolved spectra showed that the H$\delta$ EWs were also strongest at the centre, but extended over $\sim$5kpc. %Yagi Subaru
 \citet{2005ApJ...622..260S} observed a H$\delta$-strong galaxy with a weak [OII] emission  from the \citet{2003PASJ...55..771G} catalogue, and found that A-type stars are widely distributed across the system and are not centrally concentrated. Note, however, that this galaxy had a weak emission in [OII] (4.1\AA), and thus, cannot be called as an E+A galaxy in our definition. % Swinbank

It is important to keep in mind that these work has each own biases in sample selections, observations and analysis methods. 
Especially, earlier samples often lacked information on H$\alpha$, resulting in inviting contamination from H$\alpha$ emitting galaxies up to 52\% \citep[See ][]{2003PASJ...55..771G}.
 Majority of the above work, however, concluded that the post-starburst phenomena was centrally-concentrated in most cases, and significantly extended to a few kpc, being consistent with a theoretical prediction for merger/interaction remnant. 
 These results were unique in that without using information on the global/external properties of E+A galaxies, but yet reached a similar conclusion on the origin of E+A galaxies. For example, \citet{2005MNRAS.357..937G} found an excess in the number of companion galaxies of E+A galaxies, concluding that it is an evidence of the merger-interaction. Many authors reported disturbed morphologies of E+A galaxies point to the merger/interaction origin \citep{1991ApJ...381L...9O,2004MNRAS.355..713B,2006astro.ph.12053L}.

However, our results may create a stir on the interpreation of the
previous work; the E+A system J1613+5103 is cleary a dynamically-interacting
system. In \citet{YGH}, we reported a flat, uniform H$\delta$
distribution of 5 kpc.  In this work, we confirmed the result with
deeper and more extended observation, and found that the post-starburst
feature is extended to 8.5 kpc, instead of the centrally-concentrated
post-starburst region which was expected based on the merger-interaction
scenario of the E+A origin.
  We discuss possible implication of our results in the next subsection.

\begin{figure*}
\begin{center}
\includegraphics[scale=0.45]{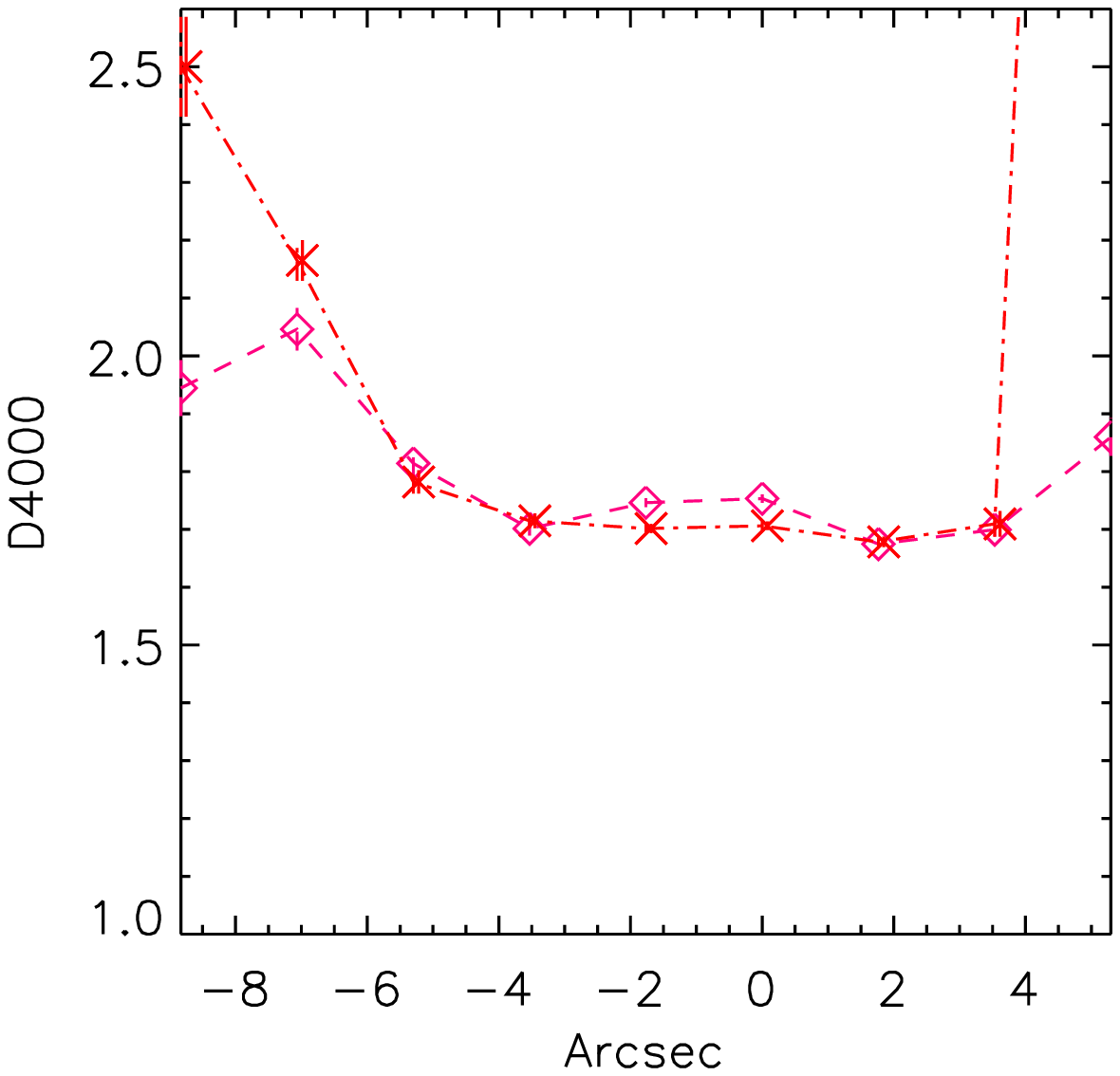}
\includegraphics[scale=0.45]{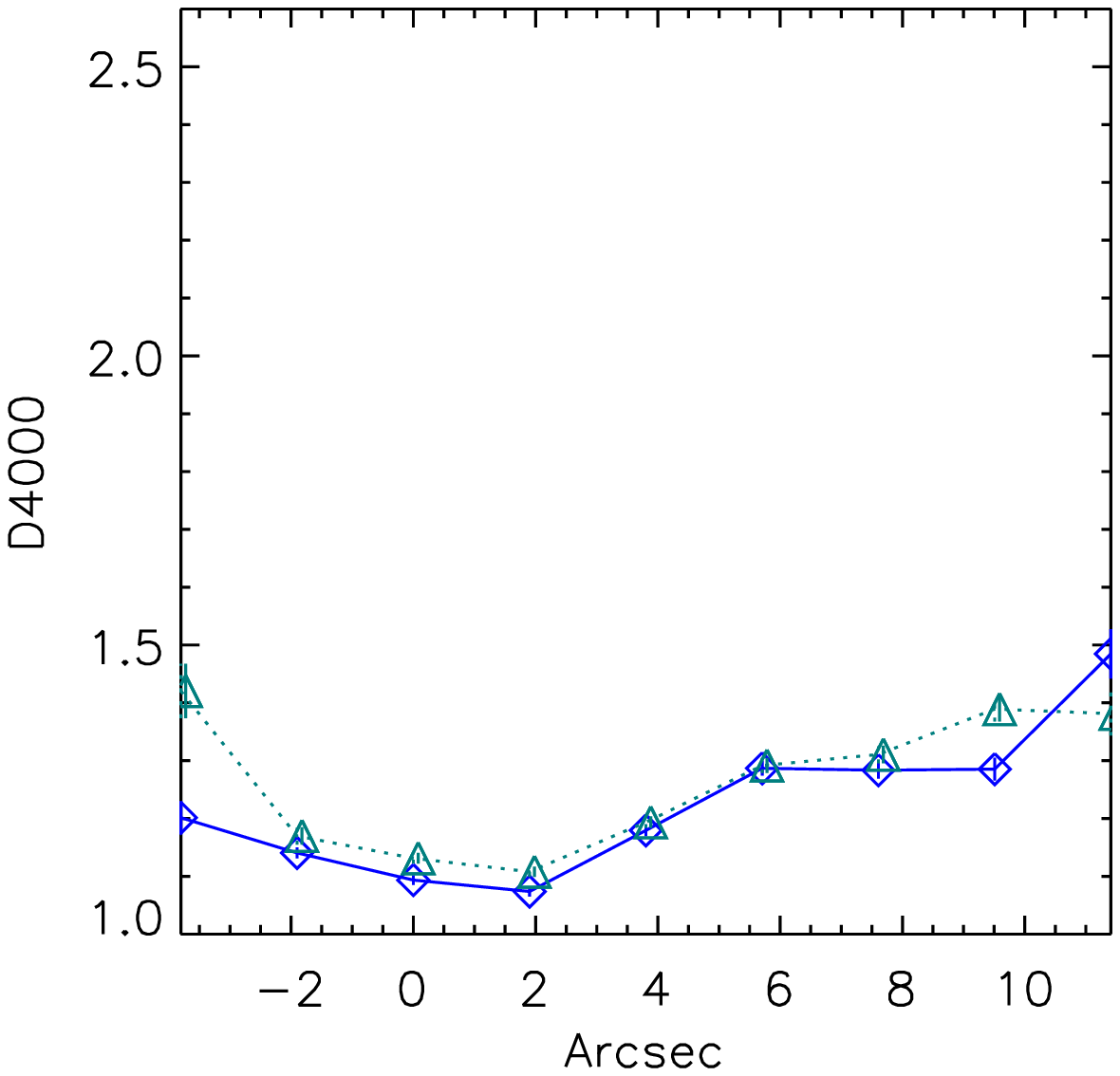}
\end{center}
\caption{The radial profile of the 4000\AA break (D4000).  
 The diamonds and triangles in the right panel are for the E+A core/north spectra, respectively. The squares and crosses are for the companion galaxy's core/north spectra in the left panel. 
 One arcsec corresponds to 0.66 kpc at this redshift.
}\label{fig:d4000_profile}
\end{figure*}

The colour gradient of this E+A galaxy was studied in detail by  \citet{2005MNRAS.359.1557Y}, who
 found that the $g-r$ and $r-i$ colours are bluer in the central region within 1.9 kpc (see their Fig.3, galaxy \#1).
 While we found a flat distribution of H$\delta$ EWs in the right panels of Fig.\ref{fig:ew}, their result again implies the existence of the age gradient; looking at the models in Fig.\ref{fig:d4000}, H$\delta$ EW does not evolve much in the first 500Myrs, especially when the burst fraction is low. In this period of $<$500Mrs, however, the D4000, which is equivalent to the colour measured out of spectra, keeps becoming larger (redder). For clarity, we plot the radial profile of the D4000 in Fig.\ref{fig:d4000_profile}, where the right panel is for the E+A core/north spectra and the left panels are for the core/north spectra of the companion. In the right panel, 
%the D4000 is smallest with $\sim$1.1 at around $x=0$ where H$\delta$ EW peaks. 
the D4000 is smaller with smaller than 1.2 at around $-2<X<4$ arcsec, which corresponds to 4 kpc wide area around the centre, in good agreement with the colour gradient of  \citet{2005MNRAS.359.1557Y}.

%gradients are  -0.173$\pm$0.667 and -0.001$\pm$0.201, i.e., almost no gradient with very large errors (see their Fig.3, galaxy \#1). Their results are consistent with our findings of the flat H$\delta$ EW profile (Fig.\ref{fig:ew}).

%a significant number of E+A galaxies exhibit a positive slope of radial colour gradient (bluer gradient toward the centre), being consistent with the hypothesis that E+A galaxies are caused by merger/interaction, having undergone a centralized violent starburst. %Yamauchi

\subsection{Physical interpretation}
 
 In this series of papers \citep[][ and this work]{YGH} we have found the following:
 Our target E+A galaxy, J1613+5103, has a companion galaxy 14 kpc away with the velocity difference of 61.8 km/s. Both the E+A and companion galaxies have disturbed morphology with the core and a fan-shaped region extended away from each other. 
 With these evidences, there is little doubt that the system is dynamically interacting galaxies. 
 Age estimate based on the H$\delta$-D4000 (Fig.\ref{fig:d4000}) or H$\delta$-[OII] (Fig.\ref{fig:hd_oii}) diagrams suggest $\sim$250 Myr after the burst, being consistent with the early stage of a dynamical interaction.
In addition, the existence of such system presents an important example that a dynamical galaxy-galaxy interaction can create an E+A galaxy as a matter of fact, and that the companion galaxy of the E+A system does not have to be an E+A galaxy. 

 In Section \ref{results}, we found that H$\delta$ EW of the E+A galaxy is very strong with $>$7\AA~ in the entire galaxy with no obvious spatial trend, despite the fact that the galaxies are dynamically interacting. 
 The results are inconsistent with the previously believed simplified picture: 
during a galaxy-galaxy interaction, the gas readily loses angular momentum due to dynamical friction, decouples from the stars, and inflows rapidly toward the merger nuclei \citep{1992ApJ...393..484B,1992ARA&A..30..705B,1996ApJ...471..115B,1994ApJ...425L..13M,1996ApJ...464..641M}, creating a central starburst, which could evolve into E+A phase if the truncation of the starburst is rapid enough \citep{2004A&A...427..125G}. In this scenario, the galaxy is left with a central population of young stars and hence a radial distribution of H$\delta$ EW which is the highest in the centre and decreases rapidly with galactocentric radius \citep{2005MNRAS.359.1421P,2005MNRAS.359..949B}.

 This scenario was the basis of the spatially-resolved observational work in the literature;
\citet{2006AJ....131.2050Y,goto_3D} found centrally-concentrated post-starburst regions (but significantly extended) in E+A galaxies, concluding it is an evidence of merger origin of the E+A galaxies.
 On the contrary, if the stripping is the cause of the E+A phenomena,  the star-formation is simultaneously and uniformly truncated throughout the entire disc, creating a flat, uniformly high $H\delta$ EW profile \citep{2005MNRAS.359.1421P}.

However, our results present an example where dynamical interaction also
can create a flat, uniform radial distribution of the post-starburst
region.
% However, our results together with \citet{YGH} presents an example where dynamical interaction also can create a flat, uniform radial distribution of the post-starburst region. 
The discovery complicates the picture; investigating a spatial distribution of the post-starburst region may not be a good way to investigate the origin of the E+A phenomena. In the case of J1613+5103, an obvious companion is present, and thus, it is easy to be judged as an interaction. However, if the minor-merger can also create such a flat profile, there is no way to distinguish from the stripping scenario. Our example cautions us that it is important to combine multiple informations together to investigate the physical origin of E+A galaxies.

 We also found that this system is likely to be an equal mass merger based on the velocity dispersion. The result is consistent with \citet{2005MNRAS.359..949B} which predicted at least a mass ratio of 0.3 is required to create an E+A galaxy through dynamical galaxy merger simulation. It is important to measure the mass of more E+A companion galaxies to constrain the mass ratio required to create an E+A galaxy. An attempt to spectroscopically follow-up E+A companions is on-going \citep{yamauchi2008}.

It is interesting that we found rotational velocity only in the companion galaxy (Fig.\ref{fig:velocity}), which did not experience a starburst. This velocity could be provoked by the dynamical interaction of the two galaxies. However, considering that these two galaxies have similar mass, and the other (E+A) does not have rotational velocity, it is more plausible that the progenitor of the companion galaxy was a rotationally supported disk galaxy. There is a hint of disk plane in Fig.\ref{fig:fchart}. Considering that the companion galaxy did not have major star-formation at least for recent 2 gigayears (Fig.\ref{fig:d4000}), the progenitor of the companion galaxy may have been a passive, rotationally-supported S0 galaxy. It also means that for at least 2 gigayears, galaxy can sustain the rotational velocity without continuing star formation activity. 
 
 But then, why E+A galaxy do not show rotational velocity even though it has experienced the starburst? 
 We have to keep in mind that the E+A could be face-on although we do not see an obvious face-on disk in Fig.\ref{fig:fchart}. 
 Since the rotational velocity of the companion galaxy is not destroyed by the dynamical interaction, it is not likely that the E+A galaxy had an initial rotation. But at the same time, E+A galaxies had experienced a starburst, therefore, must have had significant amount of gas, and possibly some star-formation activity even before the starburst. Taken all these together, one possibility, assuming not face-on, is that the progenitor of E+A galaxy in case of J1613+5103 may have been a pressure-supported, star-forming elliptical galaxy, which is quite rare in the present Universe.

 We found a weak emission in [OII] and [OIII] for both the E+A and companion galaxies.
 The finding of the [OII] emission line is apparently inconsistent with our selection criteria of E+A galaxies with no significant [OII] nor H$\alpha$ emission ([OII] EW $<2.5\AA$ and H$\alpha$ EW $<3.0\AA$). However, the detected emission is weak with EW$\sim$1\AA~ for both [OII] and [OIII]. Using the prescription for the [OII] line given in \citet{2003ApJ...599..971H}, the star formation rate of this galaxy is 0.06$M_{\odot} yr^{-1}$, which is small enough to be negligible compared with the post-starburst population.  
 Our spectra have higher resolution of $R\sim$1600 than those of the SDSS and \citet{YGH}. When we smooth the spectra to the resolution of the SDSS, the emission lines become almost invisible. This is perhaps why previous spectra did not detect the emission, and due to the same reason, we still regard the galaxy as an E+A galaxy in the post-starburst phase. 

The location of the emission lines in the E+A galaxy is puzzling. 
 The [OII] emission lines do not have enough signal-to-noise ratio to probe the location.  
 But the [OIII] emission, at the core spectrum (the upper left panel of Fig.\ref{fig:flux}), the [OIII] emission peaks at +4 arcsec away from the peak of the absorption line. The location is almost outside of the core region, and around the border of the fan-shaped extended region in Fig.\ref{fig:fchart}. However, in the north spectra (the upper right panel of Fig.\ref{fig:flux}), the [OIII] emission is peaked at the -4 arcsec, right in front of the companion galaxy.
 These locations suggest that [OIII] emission may come from outside of the post-starburst (strong Balmer absorption) core. However, the presence of only a single peak at both the core and north spectra at an opposite direction suggests that the [OIII] emission does not form a symmetric ring/sphere star-forming region, but instead it has more complicated spatial distribution.
 In this context, it will be very interesting to perform an IFU spectroscopy such as in \citet{2005ApJ...622..260S,goto_3D} for this E+A system.
 In the literature,  \citet{2005ApJ...622..260S} observed a H$\delta$ strong galaxy SDSSJ101345.39+011613.66 \citep{2003PASJ...55..771G} with the Gemini IFU, and found a similar offset of [OII] emission line from the Balmer absorptions by $\sim$2 kpc.

 For the companion galaxy, only the [OII] emission has a good enough signal-to-noise ratio to probe the spatial distribution (bottom panels of Fig.\ref{fig:flux}). The peak location of [OII] agrees with that of Balmer absorption lines, but the emission is slightly more extended to $\pm$5 arcsec on both sides, suggesting that the [OII] emission plausibly triggered by the interaction is widely distributed beyond the core region. 
The overall spectra of the companion galaxy are dominated by the old stellar population (Fig.\ref{fig:J1613_ap9}), and thus, it is not likely that the companion galaxy is experiencing a strong starburst of more than 5\% \citep{2006AJ....131.2050Y}.   Perhaps, a small amount of gas remaining in the companion galaxy was disturbed by the interaction to create a weak [OII] emission observed here. Since the emission is spatially extended, the origin of the [OII] is not likely a central AGN. 
The weak [OIII] emission at the centre also supports the lack of AGN.

\section{Conclusions}

 We have performed a spatially-resolved long-slit spectroscopy of the nearby interacting E+A system J1613+5103 using the FOCAS/Subaru. The E+A galaxy with very strong H$\delta$ EW of $\sim$7\AA~(estimated age of 250Myr) has an obvious companion at 14 kpc in front. Combined with a disturbed morphology, the system is a clear example that a dynamical interaction can actually create an E+A galaxy. 
The slit was aligned to include cores of both the E+A galaxy and the companion galaxy. Additional spectra were taken at 2 arcsec away in the north from the cores.
Our finding based on these spectra are as follows:

\begin{itemize}
 \item The strong Balmer absorption lines of the E+A galaxy are strong in the entire galaxy; Especially, H$\delta$ EW is $>$7\AA~ for the entire galaxy region ($>$8.5 kpc).  
 \item The peak of the weak (EW$\sim$1\AA) [OIII] emission of the E+A galaxy is slightly offset from that of the Balmer absorption lines, possibly suggesting that the remaining starburst is going on at a different location from the post-starburst region.
 \item For the companion galaxy emission line ([OII]) profile is more extended than that of the H$\beta$, suggesting on-going star-formation happens in more extended region than the central region of old stellar population. 
 \item We found a rotational velocity of $>$175 km/s for the companion galaxy (Fig.\ref{fig:velocity}), implying the companion was a disk galaxy without star-formation activity before the dynamical interaction. No significant rotation was found for the E+A galaxy. %, implying an interesting possibility that the progenitor of the E+A might have been a pressure-supported but yet gas-rich galaxy.
 \item Using the $r-H$ colour, we estimated the metallicity of the E+A and the companion were $Z=0.008$ and 0.02, respectively (Fig.\ref{fig:nir}). 
 \item Based on the H$\delta$-D4000 and  H$\delta$-[OII] EW plots (Fig.\ref{fig:d4000}), we estimate age of the E+A galaxy is around 250 Myr after the truncation of the burst. Possibly the centre of the E+A has a slightly younger age of $\sim$100 Myr than its outskirts.
\end{itemize}

These results present an important example that a galaxy-galaxy interaction with equal mass can create a galaxy wide, uniformly-distributed post-starburst regions. In this example, it is obvious that the dynamical interaction is the physical origin of the E+A galaxy. At the same time, it warns us that simple classification of the E+A origin with the H$\delta$ profile may not be adequate enough.

\section*{Acknowledgments}

We thank the anonymous referee for many insightful comments, which significantly improved the paper.
%We are grateful to Hisanori Furusawa for valuable help during the observation.
We are grateful to T.Hattori for valuable help during the observation.
We thank Youichi Ohyama for useful discussions. %helpful suggestions. t
%and Dr. T.Hattori for friendly help during the
%observation. 

%Dr H. C. Bhatt for a critical reading of the original version of the
%paper and an anonymous referee for very useful comments that improved
%the presentation of the paper.

T.G. acknowledges financial
support from the Japan Society for the Promotion of
Science (JSPS) through JSPS Research Fellowships for Young
Scientists.

% Use of the UH 2.2-m telescope for the observations is supported by NAOJ.
 The research was financially supported by the Sasakawa Scientific Research Grant from The Japan Science Society.
 This research was partially supported by the Japan Society for the Promotion of Science through Grant-in-Aid for Scientific Research 18840047.

The authors wish to recognize and acknowledge the very significant cultural role and reverence that the summit of Mauna Kea has always had within the indigenous Hawaiian community.  We are most fortunate to have the opportunity to conduct observations from this sacred mountain.

    Funding for the creation and distribution of the SDSS Archive has been provided by the Alfred P. Sloan Foundation, the Participating Institutions, the National Aeronautics and Space Administration, the National Science Foundation, the U.S. Department of Energy, the Japanese Monbukagakusho, and the Max Planck Society. The SDSS Web site is http://www.sdss.org/.

    The SDSS is managed by the Astrophysical Research Consortium (ARC) for the Participating Institutions. The Participating Institutions are The University of Chicago, Fermilab, the Institute for Advanced Study, the Japan Participation Group, The Johns Hopkins University, Los Alamos National Laboratory, the Max-Planck-Institute for Astronomy (MPIA), the Max-Planck-Institute for Astrophysics (MPA), New Mexico State University, University of Pittsburgh, Princeton University, the United States Naval Observatory, and the University of Washington.

%\clearpage

%\appendix

%\bsp

\label{lastpage}

\end{document}